\documentstyle[12pt]{article}
\font \bigrm = cmr17 at 24pt
\font \secfont = cmbx12 at 16pt
\def \MSbar {\vbox{\hrule\kern 1pt\hbox{\rm MS}}}
\def \GeV { {\ \rm GeV} }

\def \DESsection #1 {\bigskip\leftline{\secfont #1}\par\medskip
\noindent}
\begin{document}
\null
\rightline{hep-ph/9506218}
\vskip 4 true cm
\centerline{\bigrm Summary of the XXX Rencontre de Moriond}
\centerline{\bigrm QCD Session}
\bigskip
\centerline{Davision E.\ Soper}
\centerline{\it Institute of Theoretical Science}
\centerline{\it University of Oregon}
\centerline{\it Eugene, Oregon 97403 USA}
\vskip 6 cm
\centerline{Abstract}
The main topics covered in this summary talk are the Large Hadron
Collider, heavy ion collisions, renormalons, parton distribution
functions, measurements of the strong coupling and the top quark. Of
special interest this year was the discovery of the top quark by the
CDF and D0 groups at Fermilab. I conclude that the evidence is
compelling and that the best fit of the Standard Model top hypothesis
to the data gives a top mass of $170 \pm 9 \GeV$ with a good $\chi^2$.

\pagebreak
%
\DESsection{Introduction}
This session of the Rencontre de Moriond was packed with exciting
new physics results, which were presented in approximately 110
talks. In addition, there were lively discussions after each talk.
It is, of course, hopeless to summarize all of this in a one hour
talk. Indeed, most of the talks were themselves summaries. What I
try to do is to outline some of the main themes. In addition,
I supplement what is contained in the individual talks by
providing some commentary on particular subjects, particularly points
raised in discussions, results that some participants found
surprising, or ideas that seemed to thread through several talks. I
also present some suggestions for future theoretical efforts.

In this summary, I refer to the talks contained in these proceedings
by simply giving the author's name in italics. References to the
original literature are mostly provided in the individual talks.

Of course, I must omit many topics. For example, I am
excited by the results of the H1 and Zeus collaborations at HERA
concerning the diffractive structure function $F_2^{\rm diff}$ in
deeply inelastic scattering. As outlined in the talk of {\it
T.~Doeker}, these groups have measured deeply inelastic scattering in
events where the incoming proton is diffractively scattered,
exchanging a Pomeron with the rest of the system. (The scattered
proton is not detected in the experiments reported, but one observes
that the forward moving debris that normally results from the breakup
of the proton is absent in these events, leaving a ``rapidity gap.'')
Essentially, this is the Rutherford experiment with the Pomeron as
the target, and one finds that the Pomeron is made of pointlike
partons.

I should also note that the recent improved measurement of the mass
of the $W$ boson was covered in the electroweak session of the
Rencontre de Moriond.

\DESsection{Prospects for the Large Hadron Collider}
The Large Hadron Collider was approved this year, providing a path
for experimental investigation of physics on the TeV scale, and,
in particular, of the physics of electroweak symmetry breaking. As
reviewed in a talk by {\it J.~Aubert}, there are two general
purpose detectors planned, CMS and Atlas, plus a detector, Alice,
for investigation of heavy ion collisions. The design and prototype
testing for these detectors is underway.

The prospects for finding the Standard Model Higgs boson (if that is
what generates electroweak symmetry breaking) were reviewed by {\it
L.~Poggioli}. He reported on detailed studies with full detector
simulation. If the Higgs mass is below about 150 GeV, the strategy is
to look for its decay into two photons, and the search will be
difficult. For a heavier Higgs, one will look for decay to two gauge
bosons which in turn decay into four leptons. Here the search is
easier. For a Higgs mass above some 700 GeV, the Higgs boson is a
very broad resonance and its production cross section is small. Here,
one will have to look for decays to two leptons and two hadronic jets.
The prospects for finding supersymmetric Higgs bosons were discussed
by {\it A.~Nikitemko} and {\it D.~Graudenz}.

\pagebreak
\DESsection{Heavy Ion Collisions}
Results from heavy ion collisions were reported by {\it I.~Tserruya},
{\it G.~Paic}, {\it T.~Alber} and {\it L.~Gutay}.

The eventual goal is to investigate the quark-gluon plasma that may
form in high energy, head-on collisions of large nuclei. This is,
however, not so easy, since one must have a good probe of the plasma
dynamics, but the particles that one can observe in the final state
mostly arise from material that has cooled and returned to being
ordinary hadronic matter.

An analogy may be useful. There once was (we believe) a quark-gluon
plasma, and it filled the entire universe. If this quark-gluon plasma
had emitted weakly interacting probe particles that our eyes could
see, then we could see it by looking at the night sky. Of course,
this plasma disappeared some twelve billion years ago, but our
imaginary probe particles would still bring us the evidence after a
twelve billion year trip. Indeed, if we look at the night sky with
detectors of microwave photons, we do see a plasma. It is, however, an
electron-nucleus plasma, not the quark-gluon plasma. The photons are
left over from the time that the electrons combined with the nuclei
to form atoms. In a sense, we can also observe the hadron gas that
existed in a certain era between that of the quark-gluon plasma and
the electron-nucleus plasma. That is because the ratio of $H$ to $He$
nuclei was frozen during a certain minute of the dynamics and, not
subject to the deadening laws of equilibrium thermodynamics, has been
largely unchanged to this day.

It may prove easier to probe a man-made quark gluon plasma than it
is to see the one made by nature twelve billion years ago. One probe
is the ratio of strange to non-strange particles, which is analogous
to the $H$ to $He$ ratio mentioned above. This was touched on in the
talk of {\it C.~Merino}. Another possible probe is provided by
$J/\psi$ and $\psi'$ particles, which melt in a quark gluon plasma,
as discussed by {\it D.~Kharzeev}. A third probe is high transverse
momentum jets, which, according to the talk of {\it S.~Peign\'e},
cannot easily penetrate a quark-gluon plasma. It remains to be seen
whether any of these probes can provide a conclusive signature for
the transient formation of a quark-gluon plasma in heavy ion
collisions.

\DESsection{B Physics}
We heard results on $B$ decays to charm as measured at LEP from {\it
D.~Koetke}, on semileptonic B decays measured by CLEO from {\it D.~
Cinabro}, on B masses and lifetimes as measured at LEP from {\it
V.~Canale}, on B mixing, lifetimes and rare decays as measured by CDF
and D0 by {\it D.~Lucchesi}, on $B$ oscillations as measured at LEP by
{\it S.~Emery} and {\it L.~Brillault}, on $B$ decay to baryons as
measured by CLEO from {\it M.~Zoeller}, on polarization in $B$ decays
to $\psi K^*$ as measured by ARGUS from {\it D.~Ressing}, on $B^*$ and
$B^{**}$ states seen at LEP from {\it S.~Schael}, on $B\pi$ and $BK$
correlations observed by OPAL from {\it C.~Sheperd-Themistocleo} and
on $\Lambda_B$ production as measured by L3 from {\it M.~Lenti}. In
addition, there were results on inclusive $b$ production in hadron
collisions as measured by CDF and D0 presented by {\it L.~Markosky}
and {\it V.~Papadimitriou}. It is certainly heartening to see so much
progress in one year. I cannot help but take note of a difference
between this year's meeting and previous Rencontres de Moriond that I
have attended: the emergence of LEP and Fermilab as B-factories with
the help of silicon vertex detectors.

Let me emphasize one anomaly among the results reported. According
to our best understanding of how field theory and QCD operates, the
lifetime of a hadron containing a single heavy quark $Q$ of mass
$m_Q$ should be given by
\begin{equation}
\tau = \tau_{\rm parton}\{
1 + {\kappa^2 \over m_Q^2} +\cdots
\},
\end{equation}
where $\tau_{\rm parton}$ is independent of which hadron is decaying
and $\kappa^2$ is a matrix element of certain operators in the
hadron state. The most important operator is related to the kinetic
energy of the light parton degrees of freedom. ({\it Cf.} the talk of
{\it A.~Vainshtein}.) The power corrections above begin at $1/m_Q^2$;
there are supposed to be no $1/m_Q$ terms, although one could worry
that effects of ``ultraviolet renormalons'' might produce a $1/m_Q$
contribution. If we believe this formula for $b$ quarks with a mass
$m_b \approx 5 \GeV$, then all hadrons containing one $b$ quark
should have approximately equal lifetimes, with differences of
order $(\kappa / m_Q)^2 \sim (0.5\GeV/5\GeV)^2\sim 0.01$. The
results from LEP comparing the two $B$ mesons are in accord with
this expectation: $\tau(B^+)/\tau(B^0) = 1.08 \pm 0.08$. However,
the result $\tau(B^+)/\tau(\Lambda_B) = 1.48 \pm 0.13$ creates a
puzzle.

Another anomaly concerns the total rate for production of $b$ quarks
in $p\bar p$ collisions, as measured by D0 and CDF. The $b$ quarks
must have a transverse momentum bigger than some value $P_{T\,\rm
min}$. The experimental rate is above the QCD prediction at
next-to-leading order for all values of $P_{T\,\rm min}$, from 6 to
40 GeV. The results were presented as being in agreement with QCD
theory, but the disagreement is typically some 30\%. It seems to me
that theorists should be able to do better than that, particularly
when a $P_T$ scale of 40 GeV is involved. I would like to suggest
that for large $P_{T\,\rm min}$ it would be useful to use a
theoretical formulation in which $P_{T\,\rm min}$ provides the hard
scale. A function $f_{b/p}(x,\mu)$ that gives the distribution of $b$
quarks in a proton and a function $d_{b/a}(x,\mu)$ that gives the
distribution of $b$ quarks in the decay of a light parton $a$ appear in
this formulation. These functions can be calculated from QCD theory,
but in a calculation with a hard scale that is only $m_b$
instead of $P_{T\,\rm min}$. It would be interesting to determine
these functions from the data.  If they differ somewhat from what
calculation says they should be, it would be both interesting and
useful for other applications to know what the discrepancy is.

A third anomaly that was discussed at this meeting concerns the
production of $J/\psi$, $\psi^\prime$ and $\Upsilon$ in $p\bar p$
collisions. The rates for these processes have been problematic for
QCD theory. A more sophisticated theory seems to help, as reported
by {\it M.~Cacciari}. However, the problems did not appear to be
completely solved. I would like to point out that the theoretical
problem facing those who attempt to calculate these decay rates is
not simple, since the heavy quark mass $m_Q$ is not the only
hard scale. The inverse size of the quarkonium wave function $\sim
\alpha_s m_Q$ also enters the problem, and this scale is not so hard.
In my estimation, this field is making theoretical progress, and one
should not be discouraged if it is not completely sorted out yet.

Finally, the CLEO Collaboration reported the observation of the
decay $B \to \pi\,e\,\nu$. This decay involves a quark decay $b \to
u$. One would like to disentangle the weak from the strong
interactions so that such a measurement of a weak decay of a hadron
gives the value of the weak mixing matrix element $V_{ub}$.  A good
way to do this is to use lattice QCD to calculate the hadronic part.
For this purpose, the decay $B \to \pi\,e\,\nu$ has a special
significance, since the lattice method is best adapted to use in an
exclusive decay with a simple final state. Transitions $b \to u$ have
been observed before, but this is the first exclusive $b\to u$ decay
that has been measured.

\DESsection{Renormalons}
I now turn to a theoretical subject. Renormalons were discussed in
the talks of {\it V.~Braun} and {\it G.~Korchemsky}, but they also
entered obliquely into other talks and into the discussions. Many
Moriond participants may have wondered what this discussion meant.

Partly, what may seem to be a shift of paradigm among QCD theorists
is only a change of jargon. In a typical QCD expansion for a physical
quantity,
\begin{equation}
R(\alpha_s(Q^2)) = 1 + A_1\left({ \alpha_s(Q^2) \over \pi}\right)
 + A_2\left({ \alpha_s(Q^2) \over \pi}\right)^2
+\cdots
+{ m^2 \over Q^2} + \cdots,
\end{equation}
one has perturbative $\alpha_s^N$ terms that fall off like logarithms
of the hard scale $Q^2$ and one has power suppressed terms that fall
off like powers of $Q^2$. Theorists used to denote the power
suppressed terms by using the obscure technical term ``higher
twist.'' The new jargon is ``renormalon terms.''

There is, of course, a technical meaning, just as there was for
``higher twist.'' The word ``renormalon'' refers the Borel transform
$\tilde R(z)$ of the physical quantity $R(\alpha_s)$ and to a
certain kind of singularity of $\tilde R(z)$ in the complex $z$
plane. There are infrared and ultraviolet renormalon singularities.
Here I consider the infrared renormalons, which are the most
dangerous.

The physical interpretation of these singularities is as simple as it
is significant. Consider a graph contributing to, say, the
conventionally normalized cross section for $e^+e^- \to {\rm
hadrons}$, $R(\alpha_s(Q^2))$, where the photon virtuality $Q^2$ is
large. Suppose that the graph contains a gluon with momentum
$k^\mu$. The contribution to such a graph from the integration region
in which $k^2$ is smaller than some hadronic mass $m^2$ is small. In
this example, it is of order $m^4/Q^4$. Thus we normally don't worry
about such contributions and we perform the integration using a
perturbative gluon propagator. However, ultimately the contributions
from this integration get out of control. In fact, if we dress
the gluon propagator with gluon loops, then at order $\alpha_s^N$
the perturbative integral is dominated by the region $k^2 \sim Q^2
e^{-N}$, giving a badly behaved contribution of order $\alpha_s^N
\times N!$. Clearly if we try to go beyond a calculation of order $N
\approx \log (Q^2/m^2)$ then the dominant integration region is just
the region $k^2 < m^2$ where perturbation theory should not apply.

Thus contributions from infrared integration regions in Feynman
graphs are connected to an $N!$ growth of the value of
certain kinds of high order graphs and to power suppressed
``infrared renormalon'' contributions to physical quantities. The
power suppressed contributions are of practical importance in the
case of $\tau$ decay. To estimate the size of the power suppressed
contributions, one replaces the propagator for the low momentum gluon
by a phenomenologically determined vacuum matrix element of a gluon
operator, using the formalism of the operator product expansion.


\DESsection{Parton Distribution Functions}
One ingredient in QCD calculations involving hadrons is the parton
distribution functions. Thus the measurement of these functions is
an important goal. In addition, the consistency of measurements in a
variety of processes and at a variety of momentum scales provides a
check on the QCD theory. {\it R.~Roberts} reviewed parton
distributions for the conference. There were talks on a variety of
measurements that bear on the determination of these functions.
{\it M.~Klein} covered the measurement of $F_2(x,Q^2)$ in deeply inelastic
scattering at HERA, which I discuss briefly below. {\it A.~Kotwal}
discussed measurements of $F_2(x,Q^2)$ on protons and deuterons by
the E665 experiment (see also the review of {\it B.~Badelek}). {\it
M.~Szleper} covered measurements by the NMC collaboration of
$F_2(x,Q^2)$ on $^{119} Sn$ and $^{12}C$. There were three talks
reporting measurements by the D0 and CDF collaborations of cross
sections in $p\bar p$ collisions that provide information on parton
distributions. {\it Q.~Fan} reported on the cross section
for $W$ production, which bears on the ratio of the number of up
quarks to the number of down quarks in the proton. These measurements
helped catalyze changes in published parton distributions last year.
{\it J.~Lamoureux} talked about the cross section for $\gamma$
production in $p\bar p$ collisions, which provides information on the
gluon distribution. Finally, {\it T.~Geld} reported cross sections for
dijet production, which also provides information on the gluon
distribution.

Before proceeding to a discussion of individual reactions, I offer
two general comments. First, it became apparent more than once during
the discussions that the particle physics community could make good
use of parton distributions that provide the best fits to the world
data for a variety of choices of $\alpha_s$.  This is easy, and has
in fact been done from time to time. The MRS group provided a set a
few years ago in which there were a variety of choices for $\alpha_s$
and the shape of the gluon distribution.\cite{MRS} Last year, the CTEQ
group provided a standard set CTEQ2M and an alternative set CTEQ2ML
in which $\alpha_s$ was set to the value determined by LEP
experiments.\cite{CTEQ} There is also a need for published parton
distributions that come with an error matrix. Then, given a
calculation of a cross section, one could assign an error attributed
to uncertainties in the parton distributions. This is not easy, and
it has not been done. The best currently available method for
assigning an error attributed to uncertainties in the parton
distributions is to try the calculation with two or three published
parton sets and take the difference in the results as an error
estimate. This is similar to estimating the size of a French mountain
valley by taking the r.m.s. dispersion in the locations of
individuals in a flock of sheep grazing in the valley.

I now turn to the measurement of $F_2(x,Q^2)$ by the H1 and Zeus
collaborations at HERA. Since HERA has a large reach toward small
$x$ with still substantial $Q^2$, the greatest interest here has
been the small $x$ region. The results of both groups appear to be in
good agreement. Before the experiments were done, there had been an
expectation that $F_2(x,Q^2)$ would rise at small $x$. This
expectation was based, on one hand, on the BFKL equation for
the variation of $F_2$ as a function of $x$. On the other hand, it
was based on the evolution equation for the variation of
$F_2$ as a function of $Q^2$: if $F_2$ were flat as a function of
$x$ at a low value of $Q^2$, it would quickly develop a slope at
larger values of $Q^2$. The experimental result is that $F_2(x,Q^2)$
does rise at small $x$, although not as fast as predicted by the
lowest order BFKL equation.

The $Q^2$ variation of $F_2(x,Q^2)$ provides information on the gluon
distribution, since in the evolution equation for $\partial
F_2/\partial Q^2$ the right hand side contains a term proportional
to the gluon distribution. When the gluon distribution is thus
determined, the normal 2-loop Altarelli-Parisi evolution equation
appears to work well, despite earlier speculations that one might see
the breakdown of this equation at small $x$. This equation can be
improved with a summation of leading $\log(1/x)$ terms, as discussed
by {\it F.~Hautmann}, {\it R.~Ball}, and {\it R.~Peschanski}.

The reaction $p\,\bar p \to \gamma\,X$ plays an important role in
determining the gluon distribution. A problem with this
determination was reported in the talks of {\it R.~Roberts} and
{\it W.~Vogelsang}. There are experiments at a variety of values of
$\sqrt s$ and in each such experiment the observed cross section
falls faster with the transverse momentum $P_T$ of the photon than is
predicted by the theory using the parton distributions that give the
best overall fit to this and other data. {\it G.~Korchemsky}
presented a solution of this problem based on smearing the $P_T$
distribution of the photon\cite{CTEQphoton} with transverse momentum
generated by the emission of multiple soft gluons from the incoming
partons.

{\it P.~Grenier} and {\it J.~Saborido} presented results concerning
polarized parton distributions. These parton distributions, as
measured by the structure function $g_1(x,Q^2)$, should obey a
certain sum rule due to Bjorken. Analyses of data from experiments
E143 and E142 at SLAC and from the SMC collaboration indicate that
the Bjorken sum rule holds within the experimental errors.

Finally, {\it J.~P.~Guillet} reviewed work on parton decay functions
$d_{A/a}(z,\mu^2)$, while new data useful for their determination
were presented by {\it Y.~Yamada} (TRISTAN), {\it M.~Watson} (LEP)
and {\it K.~Baird} (SLD). The parton decay functions are just as
fundamental as the more familiar parton distribution functions
$f_{a/A}(x,\mu^2)$ --- or just as non-fundamental, depending on your
view. They are not as important as distribution functions, which are
essential for every QCD experiment with hadrons in the initial state.
Nevertheless, parton decay functions are still of substantial
practical usefulness, and in my opinion it is good that we are now
determining them from data.


%
\DESsection{Measuring the Strong Coupling}
The measurement of $\alpha_s$ is not the sole goal of QCD studies,
but it is significant as a determination of one of the
fundamental constants of nature and as an input for studies of what
may lie beyond the standard model. In addition, the agreement among
measurements made with different methods and at different scales
provides a check on the correctness of the theory. My impression from
the results presented at Moriond XXX is that the level of agreement
is not quite consistent with the expected experimental and
theoretical errors.

One needs a careful definition in order to compare results, since the
renormalization method affects the meaning of $\alpha_s$.
Fortunately, there is a consensus to adopt as a standard of
comparison $\alpha_s(\mu)$ defined in the \MSbar\ scheme with five
flavors, choosing $\mu = M_Z$. Other measurements, say measurements
of $\alpha_s(\mu)$ with three flavors at $\mu = M_\tau$, are
translated to this standard.

The talks on measurements of $\alpha_s$ were reviewed by {\it S.~Betkhe}.
He reported a value $\alpha_s(M_Z) = 0.117\pm 0.006$ based on the
previous world average together with the results reported at this
meeting. I review below a few of the new results.

The measurable quantity associated with the smallest theoretical
errors on $\alpha_s$ is the width for $Z \to {\rm hadrons}$ at LEP.
The perturbative expansion is known to next to next to leading order,
the perturbative coefficients appear to be well behaved, and the
power suppressed corrections are negligible. Unfortunately, the
theoretical expression has the form $\Gamma = \Gamma_0\{ 1 +
(\alpha_s/\pi) + \cdots\}$, so that experimental errors in the
determination of $\Gamma$ are magnified when expressed as errors on
$\alpha_s$: $(\delta\alpha_s/\alpha_s) \approx(\pi/\alpha_s)\times
(\delta\Lambda /\Lambda) \approx 30 (\delta\Lambda /\Lambda)$.
Despite the difficulties, there has been progress over the years. The
value reported twelve years ago at the 1983 Multiparticle conference
was\cite{Wolf} $\alpha_s(M_Z) = 0.153 \pm 0.050$. The LEP result
reported by {\it  J.~Casaus} at this conference was $\alpha_s(M_Z) =
0.127 \pm 0.006$. Thus the error has been reduced by an order of
magnitude in twelve years. Note that the value obtained is a bit high
compared to the world average value.

Results were also reported for event shapes in $e^+e^- \to {\rm
hadrons}$. Here the purely experimental errors are small, but the
theoretical error is large: there are only two terms known in the
perturbative expansion, the indications based on scale dependence are
that the perturbative expansion is not so well behaved, and estimated
power suppressed corrections are substantial. Still, we are better
off than we were in 1983, when the state of the theory was such
that values that were not consistent with one another were obtained.
For instance\cite{Wolf}, one analysis gave  $\alpha_s(M_Z) = 0..165
\pm 0.010$ while another gave $\alpha_s(M_Z) = 0.119 \pm 0.010$. At
this meeting, {\it K.~Baird} reported results from SLD of
$\alpha_s(M_Z) = 0.120 \pm 0.008$ and {\it J.~Casaus} reported
results from LEP of $\alpha_s(M_Z) = 0.123 \pm 0.006$. Within the
errors, these are consistent with the world average.

A measured quantity that is closely related to $\Gamma(Z \to {\rm
hadrons})$ is the width for $\tau \to \nu + {\rm hadrons}$. The
difference here is that the scale is $M_\tau$ instead of $M_Z$.
Since $M_\tau$ is so small, one wonders whether measurements of
$\Gamma(Z \to {\rm hadrons})$ can provide a credible measurement of
$\alpha_s$. In particular, the power suppressed corrections are not
negligible. The indications from careful analyses is
that the measurement is credible\cite{Braaten}, but some
authors\cite{Schifman} believe the theoretical error on
$\alpha_s(M_Z)$ is as small as $\pm 0.002$, while
others\cite{Altarelli} believe that it is no smaller than $\pm
0.006$. In my opinion the distinction is between theorists who quote
a ``1 $\sigma$'' error, with the meaning that they expect that the
error is not much smaller than the estimate given, and theorists who
quote a ``95\% confidence limit,'' with the meaning that they are
pretty sure that the error is smaller than the estimate given. It
seems to me that $\pm 0.003$ is a reasonable estimate as a 1 $\sigma$
error, while a 95\% confidence limit might be $\pm 0.010$. The
LEP values presented at this conference by {\it P.~Reeves} were
$\alpha_s(M_Z) = 0.123 \pm 0.003$  from OPAL and $\alpha_s(M_Z) =
0.122 \pm 0.003$  from ALEPH. {\it D.~Dumas} reported the result
$\alpha_s(M_Z) = 0.114
\pm 0.003$ from CLEO. {\it P.~Raczka} presented a theoretical
analysis based on previous data that gave a value $\alpha_s(M_Z) =
0.120 \pm 0.003$. It is not clear to me where the differences among
these results come from.

The value of $\alpha_s$ can also be extracted from deeply inelastic
lepton scattering.  Global fits to parton distributions produce a
fitted value of $\alpha_s$, but it is difficult to determine the
corresponding error. An evaluation based on the QCD corrections to
the Gross-Llewellyn Smith sum rule and data from the CCFR
collaboration was presented by {\it D.~Harris}: $\alpha_s(M_Z) = 0.107
^{+0.007}_{-0.009}$. Note that this result is below the world
average value.

The final example that I will discuss is based on the level
splittings of $c \bar c$ and $b\bar b$ states, which can be very
well measured. The idea is to calculate these splittings with lattice
QCD and adjust the lattice $\alpha_s$ to match the observed
splittings. One must correct for the facts that the lattice spacing
is not zero and the lattice size is not infinite. In previous years
this calculation was performed with the number of light quark
flavors set to zero, and one had to correct for the $N_f = 0$
approximation. This year there are new results with $N_f = 2$, so
this correction is smaller. Finally, one must relate the five flavor,
\MSbar\ version of $\alpha_s$ at $\mu = M_Z$ to the version of
$\alpha_s$ used on the lattice. This requires a perturbative
calculation. The result\cite{NRQCD} reported by {\it P.~Mackenzie}
is
$\alpha_s(M_Z) = 0.115 \pm 0.002$. This is a new result and may take
a year to settle down, but it appears to me that this method will set
the state of the art in reliability and precision.

\DESsection{Discovery of the Top Quark}

It was been a great joy to hear from the CDF and D0 collaborations
the details of their discovery of the long awaited top quark.  The
information was presented in talks by {\it A.~Yagil} and {\it
J.~Thompson}.

I note first that top quark production is a short distance process,
which should be calculable in perturbative QCD. For instance in the
Born graph for ${\rm gluon} + {\rm gluon} \to t + \bar t$, there is a
virtual top quark exchanged.  The virtuality $|k^2 - M_t^2|$ of this
line is of order $M_t^2$ or larger, so the virtual top quark is far
off shell. We thus expect the cross section calculated at next to
leading order to be quite accurate.

The two experimental groups used a number of different methods for
tagging events as candidates for top quark events. I provide here a
quick overview, with the warning that the reader should consult the
full talks for a more complete description.

First, both groups looked for decays of the produced $t\bar t$ into
two leptons plus jets. The idea is that each $t$ decays to $Wb$ and
the $W$ decays into a charged lepton plus a neutrino. (I group here
the dilepton analysis of CDF and the  $e\mu + {\rm jets}$, $ee + {\rm
jets}$ and $\mu\mu + {\rm jets}$ channels of D0). I show below for
each detector the expected background, the expected signal plus
background based on a 170 GeV top quark, and the observed number of
events.
\begin{center}
$\ell\,\ell + {\rm jets}$\par
\begin{tabular}{lccc}
&background&top+bkg&observed\\
CDF   & 1.3 & 4.4 & 7 \\
D0    & 0.7 & 2.3 & 3 \\
\end{tabular}
\end{center}
Evidently, the hypothesis that there is a 170 GeV standard model top
quark fits the data much better than does the hypothesis that there
is only background. Next, both groups looked for decays of the
$t\bar t$ into a single isolated lepton plus jets where one of the
jets contained a muon, presumably from a $b$ quark decay. (I group
here the SLT analysis of CDF and the  $e + {\rm jets}/\mu$ and $\mu +
{\rm jets}/\mu$ channels of D0.) The results were
\begin{center}
$\ell + {\rm jets}/\ell$\par
\begin{tabular}{lccc}
&background&top+bkg&observed\\
CDF   & 15 & 22 & 23 \\
D0    & 1.2 & 3.5 & 6 \\
\end{tabular}
\end{center}
Again, the results favor the top quark hypothesis. The D0 group also
looked for a single charged lepton plus jets without another muon to
tag a $b$ quark decay, but with more stringent requirements on the
jets. (I group here the  $e + {\rm jets}$ and $\mu + {\rm jets}$
channels of D0.) The results were
\begin{center}
$\ell + {\rm jets}$\par
\begin{tabular}{lccc}
&background&top+bkg&observed\\
D0    & 1.9 & 6.4 & 8 \\
\end{tabular}
\end{center}
Again, the top hypothesis is favored. Finally, CDF looked for
events with a lepton plus jets in which they could find the
secondary vertex from the b quark decay in their silicon vertex
detector. The results were
\begin{center}
$\ell + {\rm jets}$\ \ [secondary vertex]\par
\begin{tabular}{lccc}
&background&top+bkg&observed\\
CDF    & 7 &  24 & 27 \\
\end{tabular}
\end{center}
Thus the top quark hypothesis is favored in several different methods
of analysis. Furthermore, the expected number of events matches the
observed number pretty well, although there is a tendency for the
observed number to be greater than the expected number.

Both experimental groups give a top quark mass analysis, for which I
refer the reader to the groups' talks.

It remains for future experimental work to test whether the object
that is seen is precisely the top quark of the standard model and
not some variant of that particle. We will want to pin down, for
instance, the angular distribution with which top quarks are
produced, the branching ratios for its various decay modes, and the
momentum distributions of its decay products.

For now, however, I would like to use the information at hand to
address the question of the top quark mass. I will assume that the
D0 collaboration is right, within its stated errors.  I will
assume that the CDF collaboration is right, within its stated
errors. I will assume that the object found is indeed the
standard model top quark. And I will assume that the QCD theory that
predicts its production cross section\cite{Smith} is right, within
errors.  The question then is, what is the top quark mass and what is
its production cross section?

The CDF Collaboration quotes a mass of $(176 \pm 12.8)\ {\rm GeV}$,
where I have combined the statistical and systemic errors. They
quote a cross section of $6.8\ {+3.6 \atop -2.4}\ {\rm pb}$. Thus if
$m$ is the mass in GeV and $\sigma$ is the cross section in pb, I
assign a $\chi^2$ from the CDF measurement of
\begin{equation}
\chi^2_{\rm CDF} =
\left({m - 176 \over 12.8}\right)^2 +
\left({\log(\sigma/6.8) \over 0.307}\right)^2.
\end{equation}
(I rather arbitrarily take the errors to be Gaussian in the
logarithm of the cross section, rather than the cross section
itself.)

The D0 Collaboration quotes a mass of $(199 \pm 29.7)\ {\rm GeV}$,
where I have again combined the statistical and systemic errors. D0
shows a curve, with an error band, for the cross section as a function
of the mass of the top quark. Reading from their figure, I find that
in the region near $m = 175$ their central value for the cross
section is $\log(\sigma) \approx \log(8.7) - 0.0117 (m - 175)$, with
an error on $\log \sigma$ of about 0.338. Thus I assign a $\chi^2$
from the D0 measurement of
\begin{equation}
\chi^2_{\rm D0} =
\left({m - 199 \over 29.7}\right)^2 +
\left({\log(\sigma/8.7) + 0.0117(m - 175)\over 0.338}\right)^2.
\end{equation}

Finally, I read the theoretical cross section from the curve shown
in the D0 paper as $\log(\sigma) \approx \log(5.0) - 0.0322 (m -
175)$. I rather arbitrarily assign a 20\% error to this, based on
the belief that the parton distributions that go into the calculation
are not known to better than 10\% and the cross section is
proportional to the products of two parton distributions. This
amounts to an error of $\log(1.2) = 0.182$ on $\log \sigma$. Thus I
assign a $\chi^2$ from the theoretical prediction of
\begin{equation}
\chi^2_{\rm T} =
\left({\log(\sigma/5.0) + 0.0322(m - 175)\over 0.182}\right)^2.
\end{equation}

It is now a simple matter to choose $m$ and $\sigma$ so as to
minimize the total $\chi^2$. The minimum $\chi^2$ is
\begin{equation}
\chi^2_{\rm min} = 2.5\,.
\end{equation}
This is a very reasonable value for three degrees of freedom (five
contributions to $\chi^2$ minus two parameters fit.) The individual
contributions are $\chi^2_{\rm CDF} = 0.2$, $\chi^2_{\rm D0} =
2.0$,  and $\chi^2_{\rm T} = 0.3$.

The fitted value for the top quark mass is
\begin{equation}
m = (170 \pm 9)\ {\rm GeV},
\end{equation}
while the cross section is
\begin{equation}
\sigma = 6.5\ {+1.9 \atop -1.5}\ {\rm pb}.
\end{equation}
The mass value is lower than that quoted by either CDF or D0. The
reason is that the mass is partly determined by matching the
observed cross section to the theoretical cross section, which
decreases rather sharply with increasing mass.

\DESsection{Acknowledgements}
In preparing this talk, I benefitted from conversations with several
of the Moriond participants. I would particularly like to thank
P.~Mackenzie, G.~Martinelli and L.~McLerran for valuable insights.

On behalf of all of the participants, I would like to thank the
program committee for its work in putting together a very
interesting program, the Moriond committee and staff for an
excellent job with the arrangements, and J.~Tr\^an Thanh V\^an for
providing the overall organization. This was the thirtieth Rencontre
de Moriond. The Moriond conferences have a special quality of
presenting the latest results with a mix of participants of all ages
and an atmosphere of energetic discussion. I hope that the series
will continue for a long time.


%
\end{document}